# IMAGE2POINTS: A 3D POINT-BASED CONTEXT CLUSTERS GAN FOR HIGH-QUALITY PET IMAGE RECONSTRUCTION

*Jiaqi Cui*[1], *Yan Wang*[1*], *Lu Wen*[1], *Pinxian Zeng*[1], *Xi Wu*[2], *Jiliu Zhou*[1], *Dinggang Shen*[3,4]

[1]School of Computer Science, Sichuan University, China.
[2]School of Computer Science, Chengdu University of Information Technology, China.
[3]School of Biomedical Engineering, ShanghaiTech University, China
[4]Department of Research and Development, Shanghai United Imaging Intelligence Co., Ltd., Shanghai, China
[*]Corresponding author, email: wangyanscu@hotmail.com

## ABSTRACT

To obtain high-quality Positron emission tomography (PET) images while minimizing radiation exposure, numerous methods have been proposed to reconstruct standard-dose PET (SPET) images from the corresponding low-dose PET (LPET) images. However, these methods heavily rely on voxel-based representations, which fall short of adequately accounting for the precise structure and fine-grained context, leading to compromised reconstruction. In this paper, we propose a 3D point-based context clusters GAN, namely PCC-GAN, to reconstruct high-quality SPET images from LPET. Specifically, inspired by the geometric representation power of points, we resort to a point-based representation to enhance the explicit expression of the image structure, thus facilitating the reconstruction with finer details. Moreover, a context clustering strategy is applied to explore the contextual relationships among points, which mitigates the ambiguities of small structures in the reconstructed images. Experiments on both clinical and phantom datasets demonstrate that our PCC-GAN outperforms the state-of-the-art reconstruction methods qualitatively and quantitatively. Code is available at https://github.com/gluucose/PCCGAN.

## 1. INTRODUCTION

Positron emission tomography (PET) is an ultra-sensitive and non-invasive nuclear imaging technique and has been extensively applied in hospitals for disease diagnosis and intervention [1, 2]. In clinic, standard-dose scanning is preferred by physicians for its diagnostic quality. However, the associated high radiation of radioactive tracers inevitably causes potential hazards to the human body. On the other hand, reducing the tracer dose may induce unintended noise and artifacts, resulting in degraded image quality. One solution for this clinical dilemma is to reconstruct standard-dose PET (SPET) images from the corresponding low-dose PET (LPET) images, thereby obtaining clinically acceptable PET images while reducing radiation exposure.

In the past decade, deep learning has made a quantum leap in medical images [3-7]. Along the research direction of PET reconstruction [8-18], Gong *et al.* [9] incorporated residual convolutional neural networks (CNNs) into an iterative reconstruction framework to recover SPET images from LPET images. After that, Wang *et al.* [10] proposed a 3D conditional generative adversarial network (GAN) for estimating SPET images. Recently, Luo *et al.* [14] introduced transformer into CNN for SPET reconstruction.

Despite achieving remarkable performance, these methods which heavily rely on a voxel grid representation, wherein a 3D PET image is constructed as a stack of individual voxel "slices", face the following limitations. First, due to the intricate structure of PET images, such a voxel-based representation may struggle to preserve explicit structural information (e.g., the size and boundary of each organ or tissue) [19, 20], resulting in imprecise details (e.g., edges) in reconstructed images. Second, due to the inflexibility of the regular 3D voxel grid, current methods may insufficiently account for the fine-grained contexts (e.g., relationships and interactions among different organs and tissues), leading to ambiguities and even loss of small-sized structures.

In this paper, to resolve the first limitation above, we draw inspiration from the intrinsic geometric representation power of points, thereby representing an image in the form of points to facilitate the exact expression of structural information. As for the second limitation, inspired by [21], we develop a context clustering strategy to exploit fined-grained contextual relationships of the image based on contextual affinities among points. Overall, we propose a 3D point-based context clusters GAN, namely PCC-GAN, that utilizes point-based representation and context clustering strategy to reconstruct clinically approved SPET images from LPET images.

The contributions of this paper are summarized as follows: (1) To obtain high-quality PET images while reducing radiation exposure, we propose a novel PCC-GAN model for reconstructing SPET images from the LPET images. (2) We adopt a point-based representation to explicitly preserve the intricate structure of 3D PET images, thereby enhancing the reconstruction with sharper edge details. (3) We devise a context clustering strategy to comprehensively explore the fine-grained contexts of the image, thereby alleviating the potential ambiguities and loss of small-sized organs and tissues in the reconstructed images. (4) Experimental results demonstrate that our proposed PCC-GAN outperforms other state-of-the-art methods qualitatively and quantitatively.

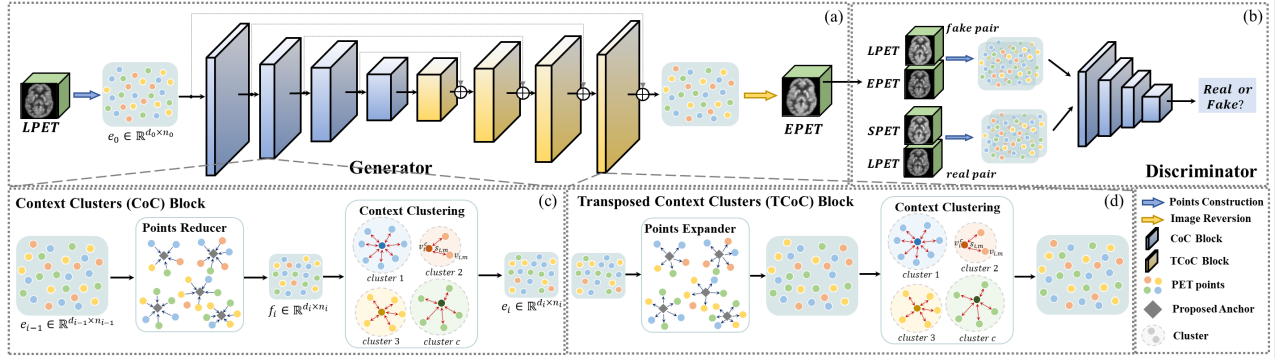

**Fig. 1.** Overview of the proposed PCC-GAN.

## 2. METHODOLOGY

The overview of our PCC-GAN is illustrated in Fig. 1, which consists of a hierarchical generator and a point-based discriminator. Specifically, taking the LPET image as input, the generator first converts it to points via points construction, and then respectively employs four context clusters (CoC) blocks and four transposed context clusters (TCoC) blocks to generate the residual points between LPET and SPET. Next, these residual points are added to the LPET points to produce the predicted PET points. After that, the predicted points are reverted to the image via points reversion, thus obtaining the output of the generator, i.e., the estimated PET image (denoted as EPET). Finally, the point-based discriminator takes the real/fake PET image pair as inputs, and determines its authenticity from the perspective of points.

### 2.1. Points Construction

Considering the disparity between the image and the points, we first transform the input 3D LPET image into set of points via points construction. Herein, we denote an LPET image $x \in \mathbb{R}^{C \times H \times W \times D}$, where $C$ represents the channel number, $H$, $W$ and $D$ are the height, width and depth, respectively. Specifically, $x$ is first converted to a point set $e_p \in R^{C \times n}$ which contains $n$ ($n = W \times H \times D$) points. Subsequently, to incorporate explicit structural information, we concatenate the 3D geometric coordinates of the points, i.e., $e_c \in R^{3 \times n}$, to $e_p$ along the channel dimension, thus obtaining the corresponding point set of the input LPET image, denoted as $e_0 = \{e_p, e_c\} \in R^{d_0 \times n_0}$, where $d_0 = C + 3$ is the point dimension. In this manner, each point contains not only the original features (e.g., texture and edge) but also explicit geometric structural information. The obtained $e_0$ is further sent to CoC blocks to excavate the contextual relationships.

### 2.2. CoC Block

#### 2.2.1. Points Reducer

To reduce computational overhead while facilitating the use of multi-scale information, we introduce a points reducer at the outset of each CoC block to decrease the number of points. As shown in Fig. 1 (c), for the $i$-th ($i$=1, 2, 3, 4) CoC block, the points reducer takes the output of the previous block $e_{i-1} \in \mathbb{R}^{d_{i-1} \times n_{i-1}}$ as input, and evenly selects $A$ ($A$=32, 16, 8, 4) anchors in the point space. Then, for each anchor, its $k$ nearest neighbor points are selected, concatenated along the channel dimension, and fused by a linear projection. Finally, we obtain a new point set $f_i \in \mathbb{R}^{d_i \times n_i}$, the point number of which is the same as the number of anchors (i.e., $A$). In this way, the point number is eighthed while the dimension is doubled layer by layer.

#### 2.2.2. Context Clustering

**Clusters Generating:** Given the point set $f_i$, we group all its points based on contextual affinities. Concretely, following the conventional SuperPixel method SLIC [22], we first evenly propose $c$ centers in the point space of $f_i$ and calculate the pair-wise cosine similarity between each point in $f_i$ and all proposed centers. Subsequently, we assign each point in $f_i$ to its most contextually similar center, thus obtaining $c$ clusters. Note that, since each point contains both original features and the geometric structural knowledge, the similarity computation accentuates not only the contextual affinities but also structural localities, thereby promoting the exploration of both contextual and structural relationships.

**Points Aggregating:** To further emphasize the contextual relationships, we dynamically aggregate all points in each cluster based on their contextual affinities to the center of the cluster. Assuming a cluster comprises $M$ points and denoting a center $v_i^c$ in the cluster's points space, we can represent the points in the cluster as $V_i = \{v_{i,m}, s_{i,m}\}_{m=1}^M \in \mathbb{R}^{M \times d_i}$, where $v_{i,m}$ and $s_{i,m}$ respectively signify the $m$-th point in $V_i$ and its similarity to the center $v_i^c$. The aggregated point $g_i \in \mathbb{R}^{d_i}$ is the contextual similarity-weighted sum of the points in the cluster to their center $v_i^c$, which can be formulated by:

$$g_i = \frac{1}{C}\left(v_i^c + \sum_{m=1}^{M} sig(\alpha s_{i,m} + \beta) * v_{i,m}\right),$$
$$s.t.\ C = 1 + \sum_{m=1}^{M} sig(\alpha s_{i,m} + \beta), \quad (1)$$

where $\alpha$ and $\beta$ are learnable parameters to scale and shift the similarity. $sig(\cdot)$ signifies the sigmoid activation function, and $C$ is the normalization factor. In this way, the contextual relationships can be accurately described by aggregating each point according to contextual affinity, thus gaining compact representations with fine-grained contexts.

**Points Dispatching:** The aggregated point $g_i$ is then adaptively dispatched to each point in the cluster also guided by contextual similarity. This approach facilitates the inter-communication among points, enabling them to collectively share the structural and contextual information in the entire cluster. Particularly, for a point $v_{i,m}$, we update it by:

$$v'_{i,m} = v_{i,m} + sig(\alpha s_{i,m} + \beta) * g_i, \quad (2)$$

After these procedures, explicit structures and fine-grained contexts are efficiently explored. Finally, we obtain the output of the $i$-th block, i.e., $e_i \in \mathbb{R}^{d_i \times n_i}$.

### 2.3. TCoC Block

As illustrated in Fig.1 (d), the structure of the TCoC block closely resembles that of the CoC block. The only difference lies in that the TCoC block utilizes the points expander to increase the point number, whereas the CoC block adopts a points reducer to decrease the point number. Specifically, asymmetric to the points reducer, the points expander treats every point in the point set as an anchor. For each anchor, a linear projection layer is applied to enlarge its channel dimension by a factor of $k$. The channel-enlarged point is divided into $k$ points along the channel dimension and are then uniformly positioned around the anchor, thereby creating an expanded point set that undergoes further processing through context clustering. The introduction of the TCoC block enables points expansion and restoration. In addition, to use the complementary information extracted by CoC blocks, the residual connection which adds the output of the TCoC block to the corresponding CoC block is utilized. Finally, the last CoC block outputs residual points between LPET and SPET which are subsequently added to the LPET points $e_0$ and are further reverted to the image, thus producing the final output of the generator, i.e., EPET image.

### 2.4. Point-based Discriminator

To further enhance image quality, we incorporate a point-based discriminator to determine the authenticity of the input image pairs, as shown in Fig.1 (b). Different from previous patch-based discriminators that discriminate a 3D image in the form of voxel patches, our discriminator determines the authenticity of the image from the perspective of points. Concretely, taking a real/fake PET image pair (i.e., the LPET image and its corresponding real SPET or fake EPET image) as input, it first employs points construction to convert the images into points and then utilizes four CoC blocks to learn more discriminative structural knowledge. Finally, a sigmoid function is applied to determine whether the input is real or not. By leveraging the inherent advantages of points, our point-based network can better discern structural disparities between the real and the reconstructed images, thus providing more informative feedback to the generator.

### 2.5. Objective Function

The objective function for our PCC-GAN is comprised of: 1) an estimation error loss, and 2) an adversarial loss.

For the estimation loss, the L1 loss is applied to enforce a close resemblance between the reconstructed EPET image $G(x)$ and the real SPET image $y$ while encouraging less blurring, which can be expressed as follows:

$$L1(G) = E_{x,y}[\|y - G(x)\|_1], \quad (3)$$

Furthermore, an adversarial loss is introduced to maintain consistency in the data distributions between the real SPET and the EPET images, which is defined as follows:

$$\min_G \max_D V(G, D) = E_{x,y}[\log D(x, y)] + E_x[\log(1 - D(x, G(x)))] \quad (4)$$

Overall, the objective function can be formulated as below.

$$V_{total} = \min_G \max_D V(G, D) + \lambda L1(G). \quad (5)$$

where $\lambda$ is the hyper-parameter to balance these two terms.

## 3. EXPERIMENTS

### 3.1. Experimental Settings

**Datasets:** The **clinical dataset** contains PET images of 8 normal control (NC) subjects and 8 mild cognitive impairment (MCI) subjects. SPET images were acquired in a 12-minute period, while LPET images were obtained in a 3-minute shortened time to simulate a quarter of the standard dose. The **phantom dataset** contains 20 simulated subjects acquired from the BrainWeb database [23]. LPET images were obtained by simulating at a quarter of the normal count level. The acquired PET images in both datasets have a size of 128×128×128. We extract 729 overlapped large patches of size 64×64×64 from each 3D image. Also, to obtain a more unbiased performance evaluation, we follow the "leave-one-out cross-validation (LOOCV)" strategy during training.

**Implementation Details:** Our model is implemented by PyTorch framework and trained on an NVIDIA GeForce RTX 3090 GPU with 24GB memory. The whole network is trained for 150 epochs by Adam optimizer with a batch size of 4. The learning rate is initialized to $2 \times 10^{-4}$ for the first 50 epochs, and linearly decays to 0 for the remaining 100 epochs. The number of neighbor points (i.e., $k$) and that of clusters (i.e., $c$) are all set to 8. $\lambda$ is empirically set as 100.

**Evaluation Metrics:** Three conventional evaluation metrics, including peak signal-to-noise (PSNR), structural similarity index (SSIM), and normalized mean squared error (NMSE) are employed to for quantitative evaluation.

Table 1. Quantitative Comparison with other state-of-the-art methods on clinical dataset.

| Methods | NC subjects | | | MCI subjects | | | #Params |
|---|---|---|---|---|---|---|---|
| | PSNR | SSIM | NMSE | PSNR | SSIM | NMSE | |
| Auto-Context [8] | 23.867 | 0.982 | 0.0235 | 24.435 | 0.977 | 0.0264 | 41M |
| Hi-Net [12] | 24.014 | 0.983 | 0.0249 | 24.430 | 0.981 | 0.0265 | 13M |
| Sino-cGAN [17] | 24.149 | 0.985 | 0.0254 | 24.224 | 0.981 | 0.0269 | 39M |
| M-UNet [11] | 24.199 | 0.984 | 0.0260 | 24.685 | 0.985 | 0.0262 | **4M** |
| LCPR-Net [13] | 24.313 | 0.985 | 0.0227 | 24.607 | 0.984 | 0.0257 | 77M |
| Trans-GAN [14] | 24.818 | 0.986 | 0.0212 | 25.249 | 0.985 | 0.0231 | 76M |
| **PCC-GAN (Proposed)** | **24.890** | **0.987** | **0.0210** | **25.271** | **0.987** | **0.0228** | 12M |

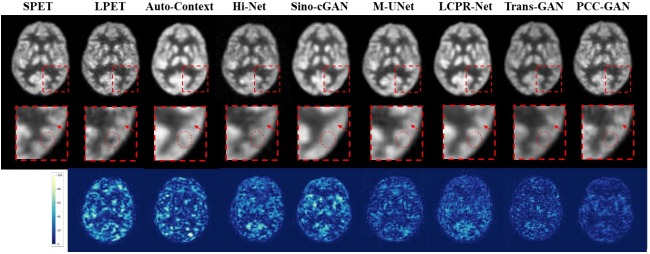

**Fig. 2.** Visualization of the reconstruction results produced by different methods on clinical dataset.

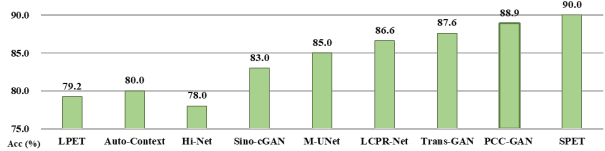

**Fig. 3.** Comparison results on clinical diagnosis experiment.

Table 2. Quantitative Comparison of our proposed method with other state-of-the-art methods on phantom dataset.

| Methods | PSNR | SSIM | NMSE |
|---|---|---|---|
| Auto-Context [8] | 27.228 | 0.984 | 0.008 |
| Hi-Net [12] | 26.467 | 0.978 | 0.011 |
| Sino-cGAN [17] | 26.768 | 0.984 | 0.010 |
| M-UNet [11] | 26.926 | 0.982 | 0.009 |
| LCPR-Net [13] | 27.123 | 0.987 | 0.008 |
| Trans-GAN [14] | 28.166 | 0.989 | 0.005 |
| **PCC-GAN** | **28.169** | **0.989** | **0.005** |

Table 3. Ablation study of key components on clinical dataset.

| Methods | NC subjects | | | MCI subjects | | |
|---|---|---|---|---|---|---|
| | PSNR | SSIM | NMSE | PSNR | SSIM | NMSE |
| (A) | 24.024 | 0.983 | 0.0231 | 24.317 | 0.981 | 0.0256 |
| (B) | 24.131 | 0.983 | 0.0229 | 24.344 | 0.982 | 0.0251 |
| (C) | 24.518 | 0.984 | 0.0217 | 24.756 | 0.984 | 0.0235 |
| (D) | 24.352 | 0.983 | 0.0220 | 24.405 | 0.983 | 0.0243 |
| **(E)** | **24.890** | **0.987** | **0.0210** | **25.271** | **0.987** | **0.0228** |

## 3.2. Experimental Results

**Comparative Results:** We compare our method with six state-of-the-art PET reconstruction approaches, including Auto-Context [8], Hi-Net [12], Sino-cGAN [17], M-UNet [11], LCPR-Net [13], and Trans-GAN [14]. The quantitative comparison results on clinical and phantom datasets are given in Table. 1 and Table. 2, respectively. It can be observed that our PCC-GAN achieves the best results among all the evaluation criteria with relatively fewer parameters on both datasets. Specifically, compared to the current best Trans-GAN, our method still boosts the PSNR by 0.072dB for NC subjects and 0.022dB for MCI subjects on the clinical dataset in Table 1. Besides, the p-values of the paired t-test between our model and the Trans-GAN are consistently less than 0.05 for both NC and MCI subjects in all metrics, verifying that our improvements are statistically significant. Moreover, our method has only 1/6 parameters of Trans-GAN, highlighting our advantage in both performance and efficiency. In addition, the visualizations in Fig. 2 illustrate that the images generated by our methods yield the best visual effect with the minimum error. All these results demonstrate the superiority of our method in predicting accurate SPET images.

**Evaluation on Clinical Diagnosis:** To prove the clinical value of our method, we further conduct an Alzheimer's disease diagnosis experiment as a downstream task on the clinical dataset. As shown in Fig. 3, the classification accuracy of our method (i.e., 88.9%) is the closest to that of SPET images (i.e., 90.0%), indicating the huge clinical potential of our method in facilitating disease diagnosis.

**Ablation Study:** To verify the contributions of the key components, we conduct the ablation study through: (A) replacing the CoC (TCoC) blocks with convolution (transposed convolution) layers, and using a patch-based discriminator, (B) replacing the patch-based discriminator with our point-based one, (C) replacing the convolution layers with CoC blocks, (D) replacing the context clustering in CoC blocks with linear layers, (E) replacing transposed convolution layers with TCoC blocks, i.e., the proposed model. As shown in Table. 3, the performance progressively improves with the introduction of each key component, thus validating their effectiveness. Moreover, comparing model (C) and (D), when replacing the context clustering with linear layers, the performance largely decreases as the model fails to learn accurate contextual information.

## 4. CONCLUSION

In this paper, we propose a 3D point-based context clusters GAN to reconstruct high-quality SPET images from LPET images. By harnessing the geometric representation of points, our method explicitly preserves the intricate structure of 3D PET images, enhancing the reconstruction with richer details. Meanwhile, we utilize a context clustering strategy to explore fine-grained contextual relationships, thereby mitigating the ambiguities or omissions of small-sized structures. Extensive experiments have demonstrated the superiority of our method.

**Acknowledgement.** This work is supported by the National Natural Science Foundation of China (NSFC 62371325, 62071314), Sichuan Science and Technology Program 2023YFG0263, 2023YFG0025, 2023NSFSC0497.